\documentstyle[12pt]{article}

\def\bkR{{\rm I\kern-.17em R}}
\def\bkC{{\rm \kern.24em \vrule width.05em height1.4ex depth-.05ex \kern-.26em C}}
\def\bkN{{\rm \kern.50em \vrule width.05em height1.4ex depth-.05ex \kern-.26em N}}

\begin{document}
\title{A global, dynamical formulation of quantum confined systems}
\author{Nuno Costa Dias\footnote{{\it ncdias@meo.pt}} \\ Jo\~{a}o Nuno Prata\footnote{{\it joao.prata@mail.telepac.pt}} \\ {\it Departamento de Matem\'atica} \\
{\it Universidade Lus\'ofona de Humanidades e Tecnologias} \\ {\it Av. Campo Grande, 376, 1749-024 Lisboa, Portugal}\\
{\it and}\\
{\it Grupo de F\'{\i}sica Matem\'atica}
{\it Universidade de Lisboa}\\
{\it Av. Prof. Gama Pinto 2,}
{\it 1649-003, Lisboa, Portugal}}
\date{}
\maketitle

{\it AMS subjclass[2000]:} Primary 47B25, 47D06; Secondary 81Q10, 34L40.

{\it Keywords:} Self-Adjoint Extensions, Boundary Interactions, Dynamical Confinement.

\begin{abstract}
A brief review of some recent results on the global self-adjoint formulation of systems with boundaries is presented. We specialize to the 1-dimensional case and obtain a dynamical formulation of quantum confinement.
\end{abstract}

\section{Introduction}

Let $H_0:{\cal D}(H_0) \subset L^2(\bkR^d) \longrightarrow L^2(\bkR^d)$ be a self adjoint (s.a.) Hamiltonian operator defined on the domain ${\cal D}(H_0)$ and describing the dynamics of a $d$-dimensional quantum system. Let us also consider the decomposition $\bkR^d=\Omega \cup \Omega^c$, where $\Omega$ is an open set and $\Gamma=\overline{\Omega}\cap\Omega^c$ is the common boundary of the two open sets $\Omega_1=\Omega$ and $\Omega_2=\overline{\Omega}^c$.

To obtain a confined version (for instance, to $\Omega_1$) of the system described by $H_0$, the standard approach is to determine the s.a. realizations of the operator $H_0$ in $L^2(\Omega_1)$. It is well known, however, that this formulation displays several inconsistencies \cite{Garbaczewski,Isham,Bonneau}, the main issues being the ambiguities besetting the physical predictions (when there are several possible self-adjoint realizations of $H_0$ in $L^2(\Omega_1)$), the lack of self-adjoint (s.a.) formulations of some important observables in $L^2(\Omega_1)$ and the difficulties in translating this approach to other (non-local) formulations of quantum mechanics, like the deformation formulation \cite{Dias1}. These problems are well illustrated by textbook examples \cite{Garbaczewski,Dias1,Akhiezer}.

Our aim here is to present an alternative approach to quantum confinement. This formulation consists in determining all s.a. Hamiltonian operators $H:{\cal D}(H) \subset L^2(\bkR^d) \longrightarrow L^2(\bkR^d)$ - defined on a dense subspace ${\cal D}(H) $ of the global Hilbert space $L^2(\bkR^d)$ - which dynamically confine the system to $\Omega_1$ (or $\Omega_2$) while reproducing the action of $H_0$ in an appropriate subdomain.
More precisely, let $ P_{\Omega_k}$ be the projector operator onto $\Omega_k$, $k=1,2$, i.e.
\begin{equation}
 P_{\Omega_k} \psi= \chi_{\Omega_k} \psi \quad , \quad  \psi \in L^2(\bkR^d)
\end{equation}
where $\chi_{\Omega_k}$ is the characteristic function of ${\Omega_k}$: $\chi_{\Omega_k}(x)=1$ if $x \in {\Omega_k}$ and $\chi_{\Omega_k} (x)=0$, otherwise. Our aim is to determine all linear operators $H:{\mathcal D}({H}) \subset L^2(\bkR^d) \to L^2(\bkR^d)$ that satisfy the following three properties:

(i) $ H$ is self-adjoint on $L^2(\bkR^d)$.

(ii) If $\psi \in {\mathcal D}(H)$ then $P_{\Omega_k} \psi \in {\mathcal D}(H)$ and $[ P_{\Omega_k}, H]\psi=0$, $k=1,2$.

(iii) $ H\psi = H_0\psi$ if $\psi \in {\mathcal D}({H_0})$ is an eigenstate of $ P_{\Omega_k}$.

Moreover, for the $1$-dimensional case, we want to recast the operators $H$ in the form $H=H_0 + B^{BC}$, where $B^{BC}$ is a distributional boundary potential (that may depend on the particular boundary conditions satisfied by the domain of $H$) and $H$ is s.a. on its maximal domain.
This formulation is global, because the system is defined in $L^2(\bkR^d)$, and the confinement is dynamical, i.e. it is a consequence of the initial state and of the Hamiltonian $H$. Indeed, from (i) and (ii) it follows that $ P_{\Omega_k}$ commutes with all the spectral projectors of $ H$ and so also with the operator $\exp \{ i  H t \}$ for $t \in \bkR$. Hence, if $\psi $ is an eigenstate of $ P_{\Omega_k}$ it will evolve to $\exp \{i  H t \}\psi $, which is again an eigenstate of $ P_{\Omega_k}$ with the same eigenvalue. In other words, $ P_{\Omega_k}$ is a constant of motion and a wave function confined to ${\Omega_1}$ (or to ${\Omega_2}$) will stay so forever.

The problem of determining a dynamical formulation of quantum confinement can be addressed from the point of view of the study of s.a. extensions of symmetric restrictions \cite{Garbaczewski,Posilicano2,Albeverio,Posilicano1} and is closely related with the subjects of point interaction Hamiltonians \cite{Albeverio,Berezin,Ph,Posilicano3} and surface interactions \cite{Kanwal}. Our results may be useful in this last context as well as for the deformation quantization of systems with boundaries \cite{Dias1}.

In this paper we shall provide a concise review of the solutions to the above problems. The reader should refer to \cite{Dias2,Dias3} for a detail presentation, including proofs of the main theorems, the extension of the boundary potential formulation to higher dimensions and some applications to particular systems.

\section{Confining Hamiltonians defined on $L^2(\bkR^d)$}

We start by introducing some relevant notation. Let $X,Y \subset V$ be two subspaces of a vector space $V$ such that $X \cap Y =\{0\}$, then their direct sum is denoted by $X \oplus Y$.
Let now $A,B$ be two linear operators with domains ${\mathcal D}(A),{\mathcal D}(B) \subset L^2(\bkR^d)$ such that ${\mathcal D}(A) \cap {\mathcal D}(B)=\{0\}$, then the operator $A\oplus B$ is defined by:
\begin{equation}
A \oplus B : \left\{ \begin{array}{l}
{\mathcal D}({A\oplus B})={\mathcal D}(A) \oplus {\mathcal D}(B)\\
=\{\psi \in L^2(\bkR^d) : \psi=\psi_1 + \psi_2, \, \psi_1\in {\mathcal D}(A), \, \psi_2 \in {\mathcal D}(B)\} \\
\\
(A\oplus B) \, \psi = A \psi_1 +B \psi_2, \, \forall \psi \in {\mathcal D}({A\oplus B})
\end{array} \right.
\end{equation}
For simplicity let us assume that ${\mathcal D}(\Omega_k) \subset L^2(\Omega_k) \cap {\mathcal D}({H_0})$, $k=1,2$  (where ${\mathcal D}(\Omega_k)$ is the space of infinitely smooth functions $t:\bkR^d \to \bkC$ with support on a compact subset of $\Omega_k$) and let us define the operators:
\begin{equation}
{H}_k^S: {\mathcal D}(\Omega_k) \longrightarrow L^2(\Omega_k), \, \phi \longrightarrow {H}_k^S \phi =  H_0 \phi \quad , \quad k=1,2
\end{equation}
which are symmetric. Let also $H_k^{S^{\dagger}}$ be the adjoint of $H_k^S$.

Our main result characterizes the operators $ H:{\mathcal D}({H}) \subset L^2(\bkR^d) \to L^2(\bkR^d)$, associated to a s.a. $H_0$, and satisfying  properties (i) to (iii). \\
\\
{\bf Theorem 1}\\
{\it Let $ H_0$ be s.a. on $L^2(\bkR^d)$ and such that ${\mathcal D}({H_0}) \supset {\mathcal D}(\Omega_1) \cup {\mathcal D}(\Omega_2)$ and $[ H_0, P_{\Omega_k}] \psi =0 $, $k=1,2$, $\forall \psi \in {\mathcal D}(\Omega_1) \cup {\mathcal D}(\Omega_2)$. An operator $ H$ satisfies the defining properties (i) to (iii) iff it can be written in the form $H_1 \oplus  H_2$ for some $ H_1, H_2$ s.a. extensions of the restrictions (3). Moreover, all operators $H$ are s.a. extensions of ${H}_1^S \oplus {H}_2^S$ and s.a. restrictions of ${H}_1^{S^{\dagger}} \oplus {H}_2^{S^{\dagger}}$.}
\\

The condition (stated in the theorem) that $[ H_0,  P_{\Omega_k}]\psi =0$, $ \forall \psi \in {\mathcal D}(\Omega_1) \cup {\mathcal D}(\Omega_2)$, and the assumption that ${H}_1^S$ and ${H}_2^S$ have s.a. extensions are the minimal requirements for the existence of operators $H$ satisfying (i) to (iii).
Proofs of these results are given in \cite{Dias2,Dias3}.

We now focus on the case where $d=1$, $\Omega_1=\bkR^-$ and
\begin{equation}
H_0=-\frac{d^2}{d x^2}+V(x) , \quad {\cal D}({H_0})= \{\psi \in L^2(\bkR): \psi,\psi' \in AC(\bkR) ;\,  H_0 \psi \in L^2(\bkR)\}
\end{equation}
where $AC(\bkR)$ is the set of absolutely continuous functions on $\bkR$ and $V(x)$ is a regular potential. We shall assume it to be  i) real, ii) locally integrable and satisfying iii) $V(x) > -kx^2, \, k>0$ for sufficiently large $|x|$.
The conditions on $V(x)$ are such that $ H_0: {\cal D}({H_0}) \subset L^2(\bkR) \to L^2(\bkR)$ is the unique s.a. realization of the differential expression $-\frac{d^2}{d x^2}+V(x)$ on $L^2(\bkR)$ \cite{Voronov} and ensure that all s.a. realizations of $-\frac{d^2}{d x^2}+V(x)$ on the semi-axes $]-\infty,0]$ and $[0,+\infty[$ are determined by boundary conditions at $x=0$ only.

For $H_0$ of the kind (4) the s.a. operators $H=H_1\oplus H_2$ are all of the form \cite{Dias2,Voronov}:
\begin{equation}
 H^{\lambda_1,\lambda_2}=  H_1^{\lambda_1} \oplus  H_2^{\lambda_2}:  \, \left\{ \begin{array}{l}
{\cal D}({H^{\lambda_1,\lambda_2}})={\cal D}({H_1^{\lambda_1}})\oplus {\cal D}({H_2^{\lambda_2}})\\
\\
 H^{\lambda_1,\lambda_2} \psi =  H^{S^{\dagger}}\psi
\end{array} \right.
\end{equation}
where
\begin{equation}
{\cal D}({H_k^{\lambda_k}})=\{ \psi_k = \chi_{\Omega_k}\phi_k :\,\phi_k \in {\cal D}(H_0) \wedge \phi'_k (0) =\lambda_k \phi_k (0) \}
\end{equation}
$\lambda_k \in \bkR \cup \{\infty \}$, $k=1,2$ and the case $\lambda_k=\infty$  corresponds to Dirichlet boundary conditions. Moreover
\begin{equation}
 H^{S^{\dagger}}= H_1^{S^{\dagger}} \oplus  H_2^{S^{\dagger}}: \, \left\{ \begin{array}{l}
{\cal D}({H^{S^{\dagger}}})=\{ \psi =\chi_{\Omega_1} \phi_1+\chi_{\Omega_2} \phi_2: \, \phi_1,\phi_2 \in {\cal D}({H_0}) \}\\
\\
 H^{S^{\dagger}}\psi = \chi_{\Omega_1}  H_0 \phi_1+ \chi_{\Omega_2}  H_0 \phi_2
\end{array} \right.
\end{equation}
Hence, all s.a. confining Hamiltonians of the form $H_1\oplus H_2$ are s.a. restrictions of $H^{S^{\dagger}}$.
To proceed let us define the operators ($k=1,2$ and $n=0,1$):
\begin{equation}
\hat{\delta}^{(n)}_k(x): {\cal D}(H^{S^{\dagger}}) \longrightarrow {\cal D}'(\bkR); \, \psi=\chi_{\Omega_1}\phi_1+\chi_{\Omega_2}\phi_2 \longrightarrow \hat{\delta}^{(n)}_k(x)\psi =\delta^{(n)} (x) \phi_k(x)
\end{equation}
where ${\cal D}'(\bkR)$ is the space of Schwartz distributions on $\bkR$ and $\delta^{(0)}(x)=\delta(x)$ and $\delta^{(1)}(x)=\delta'(x)$ are the Dirac measure and its first distributional derivative.
We can now recast the operators (5) in the additive form $H=H_0+B^{BC}$:\\
\\
{\bf Theorem 2}\\
{\it The s.a. Hamiltonian $ H^{\lambda_1,\lambda_2}$ given by eq.(5) act as:
\begin{equation}
 H^{\lambda_1,\lambda_2} \psi= \left\{ H_0 - B_1^{\lambda_1}+ B_2^{\lambda_2} \right\} \psi , \quad \forall \psi \in {\cal D}({H^{\lambda_1,\lambda_2}})
\end{equation}
where now $H_0$ is the extension to the set of distributions of the original Hamiltonian given in (4), $H_0:{\cal D}'(\bkR) \longrightarrow {\cal D}'(\bkR)$, and
\begin{equation}
{B}_{k}^{\lambda} \equiv \, \left\{ \begin{array}{l}
-\hat{\delta}_{k}'(x) +(-1)^k \hat{\delta}_{k}(x), \quad \lambda = \infty \\
\\
\hat{\delta}_{k}'(x) + 2 \lambda \hat{\delta}_{k}(x) +(-1)^k \frac{d}{d x} \left[\hat{\delta}_{k}(x) \left(\frac{d}{d x}-\lambda \right) \right] , \quad \lambda \not= \infty
\end{array} \right. \quad k=1,2
\end{equation}
Moreover, the maximal domain of the expression (9) coincides with ${\cal D}({H^{\lambda_1,\lambda_2}})$ (5), i.e.
\begin{equation}
{\cal D}_{max}(H^{\lambda_1,\lambda_2})\equiv \{\psi \in L^2(\bkR):\,
H^{\lambda_1,\lambda_2}\psi \in L^2(\bkR) \} = {\cal D}(H^{\lambda_1,\lambda_2}).
\end{equation}
}

The proof is given in \cite{Dias2}.

\section*{Acknowledgments}

We thank A. Posilicano and P. Garbaczewski for several discussions. This work was partially supported by the grants POCTI/0208/2003 and\\ PTDC/MAT/69635/2006 of the Portuguese Science Foundation.


\begin{thebibliography}{99}

\bibitem{Garbaczewski} Garbaczewski, P and Karwowski, W. Impenetrable barriers and canonical quantization. {\it Am. J. Phys.}, 2004, {\bf 72}, 924.

\bibitem{Isham} Isham, C. Topological and global aspects of quantum theory, In {\it Les Houches, Session XL}, (DeWitt, B.S. and Stora, R., eds). Elsevier, 1984.

\bibitem{Bonneau} Bonneau, G.,Faraut, J. and Valent, G. Self-adjoint extensions of operators and the teaching of quantum mechanics. {\it Am. J. Phys.}, 2001, {\bf 69}, 53.

\bibitem{Dias1} Dias, N. C. and Prata, J. N. Wigner functions with boundaries. {\it J. Math. Phys.}, 2002, {\bf 43}, 4602.

\bibitem{Akhiezer} Akhiezer, N. and Glazman, I. {\it Theory of linear operators in Hilbert space}. Pitman, Boston, 1981.

\bibitem{Posilicano2} Posilicano, A. Self-Adjoint Extensions of Restrictions. {\it Operators and Matrices}, 2008, {\bf 2}, 483.

\bibitem{Albeverio} Albeverio, S., Gesztesy, F., H\"ogh-Krohn, R. and Holden, H. {\it Solvable Models in Quantum Mechanics}, 2nd ed. AMS, Chelsea, 2005.

\bibitem{Posilicano1} Posilicano, P. A Krein-like Formula for Singular Perturbations of Self-Adjoint Operators and Applications. {\it J. Funct. Anal.}, 2001, {\bf 183}, 109.

\bibitem{Berezin} Berezin, F. and Fadeev, L. Remark on the Schr\"odinger equation with singular potential. {\it Dokl. Akad. Nauk. SSSR}, 1961, {\bf 137}, 1011.

\bibitem{Ph} Blanchard, Ph., Figari, R. and Mantile, A. Point Interaction Hamiltonians in Bounded Domains. Physics Archives: 0704.3249, 2007.

\bibitem{Posilicano3} Posilicano, A. The Schr\"odinger Equation With a Moving Point Interaction In Three Dimensions. {\it Proc. Amer. Math. Soc.}, 2007, {\bf135}, 1785.

\bibitem{Kanwal} Kanwal, R. P. {\it Generalized Functions: Theory and Technique}, 2nd ed. Birkh\"auser, Boston, 1998.

\bibitem{Dias2} Dias, N. C., Posilicano, A. and Prata, J. N. Self-adjoint, globally defined Hamiltonian operators for systems with boundaries.
Physics Archives: math-ph/0707.0948, 2007.

\bibitem{Dias3} Dias, N. C., Posilicano, A. and Prata, J. N. In preparation.

\bibitem{Voronov} Voronov, B., Gitman, D. and Tyutin, I. Self-adjoint differential operators associated with self-adjoint differential expressions. Physics Archives: quant-ph/0603187, 2006.









\end{thebibliography}
\end{document}